\documentclass{article}
\usepackage[T1]{fontenc} 
\usepackage[utf8]{inputenc} 
\usepackage{ismir,amssymb,amsmath,mathtools,cite,url}
\usepackage[USenglish,american]{babel}
\usepackage{graphicx}
\usepackage{booktabs}
\usepackage{color}
\usepackage{balance}

\usepackage{url}

\usepackage{algorithm}
\usepackage{algpseudocode}
\usepackage{makecell}

\usepackage{lineno}

\newcommand{\vect}[1]{\boldsymbol{#1}}
\newcommand{\set}[1]{\left\{ #1 \right\}}

\newcommand{\expectationwrt}[2]{\mathbb{E}_{#1}\!\left[#2\right]}

\usepackage[colorinlistoftodos]{todonotes}
\usepackage{soul}

\title{Audio-guided Album Cover Art Generation\\with Genetic Algorithms}

\multauthor
{James Marien \hspace{1cm} Sam Leroux \hspace{1cm} Bart Dhoedt \hspace{1cm} Cedric De Boom} { 
 IDLab -- Department of Information Technology, Ghent University (Belgium) -- imec\\
{\tt\small <firstname>.<lastname>@ugent.be}
}

\def\authorname{J.~Marien, S.~Leroux, B.~Dhoedt, and C.~De Boom}

\begin{document}

\maketitle
\begin{abstract}
Over 60,000 songs are released on Spotify every day, and the competition for the listener's attention is immense.
In that regard, the importance of captivating and inviting cover art cannot be underestimated, because it is deeply entangled with a song's character and the artist's identity, and remains one of the most important gateways to lead people to discover music.
However, designing cover art is a highly creative, lengthy and sometimes expensive process that can be daunting, especially for non-professional artists.
For this reason, we propose a novel deep-learning framework to generate cover art guided by audio features.
Inspired by VQGAN-CLIP, our approach is highly flexible because individual components can easily be replaced without the need for any retraining.
This paper outlines the architectural details of our models and discusses the optimization challenges that emerge from them.
More specifically, we will exploit genetic algorithms to overcome bad local minima and adversarial examples.
We find that our framework can generate suitable cover art for most genres, and that the visual features adapt themselves to audio feature changes.
Given these results, we believe that our framework paves the road for extensions and more advanced applications in audio-guided visual generation tasks.
\end{abstract}

\section{Introduction}
The invention of generative adversarial networks (GANs) by Goodfellow et al.~is commonly regarded as the dawn of large-scale, high-quality, photo-realistic image synthesis using deep generative models \cite{Goodfellow:2014td, Radford:2016, Arjovsky:2017}.
The trend continues to this day, with ever-increasing levels of detail and resolution \cite{Karras:2018, Brock:2019, Dhariwal:2021} and the ability for user-controlled image properties \cite{Karras:2019}.
Album cover art generation is a specific application of image synthesis.
Because of geometrical and convenience reasons linked to the earliest release of LPs, cover art is square-shaped and acts as a visual gateway to an album or a song.
Physical and digital browsing of music -- but also books and movies -- heavily relies upon these visual cues and has been shown to be among the most important factors in serendipitous retrieval of multimedia \cite{ross:1999, cunningham:2013, mckay:2017}.
Therefore, cover art's main function is to attract an audience, but it can also aid DJs in more efficient song retrieval.
Some fanatic listeners even cherish cover art as a true masterpiece, which explains (among others) the recent resurgent popularity of LPs.
Record labels and big artists can often afford putting a lot of work and effort into designing fitting and compelling cover art for their to-be-released song or album.
For independent and smaller artists without a strong marketing team, however, it can be a daunting and costly challenge.
For this reason, we present a GAN-powered framework that can be used to generate novel album cover art.
And because cover art is intricately connected with the songs that appear on it, our framework uses audio information as a conditioning signal to influence what the generated cover will look like.

Conditioning neural networks on extra side-information is traditionally achieved by concatenating, adding or multiplying this side-information together with the input data.
Such direct conditioning mechanisms have been generalized by the FiLM formalism \cite{dumoulin:2018}.
In a similar fashion, GANs can also be conditioned on side-information to steer image synthesis.
Most notably, C-GAN and AC-GAN both use class labels as extra input to influence the appearance of the image \cite{mirza:2014, odena:2017}, and Reed et al.~were among the first to generate images guided by textual input prompts \cite{reed:2016}.
In the context of this paper, Yang et al.~generated visual sports scenes by input-conditioning on ambient audio recordings \cite{Yang:2020}, Zhao et al.~maintained the same strategy to generate images of musicians playing instruments \cite{Zhao:2022}, and \.{Z}elaszczyk and Ma\'ndziuk leveraged variational autoencoders to model a joint audio and visual latent space \cite{zelaszczyk:2021}.
More specifically for cover art generation, Hepburn et al.~used an AC-GAN architecture to condition covers on a genre label \cite{hepburn:2017}.
The student team ``Go Deep or Go Home'' went one step further and used an architecture similar to Reed et al.'s \cite{reed:2016} to condition on learned audio features \cite{go_deep_or_go_home:2020}.
Now, although input-conditioning has been effectively applied in the works mentioned above, its main issue is that retraining is needed when we want to condition the generated images on new side-information or data modalities.
Since training GANs is a costly, time-consuming and resource-intensive process, this is to be avoided as much as possible.

\begin{figure*}[t!]
    \centering
    \includegraphics[width=\linewidth]{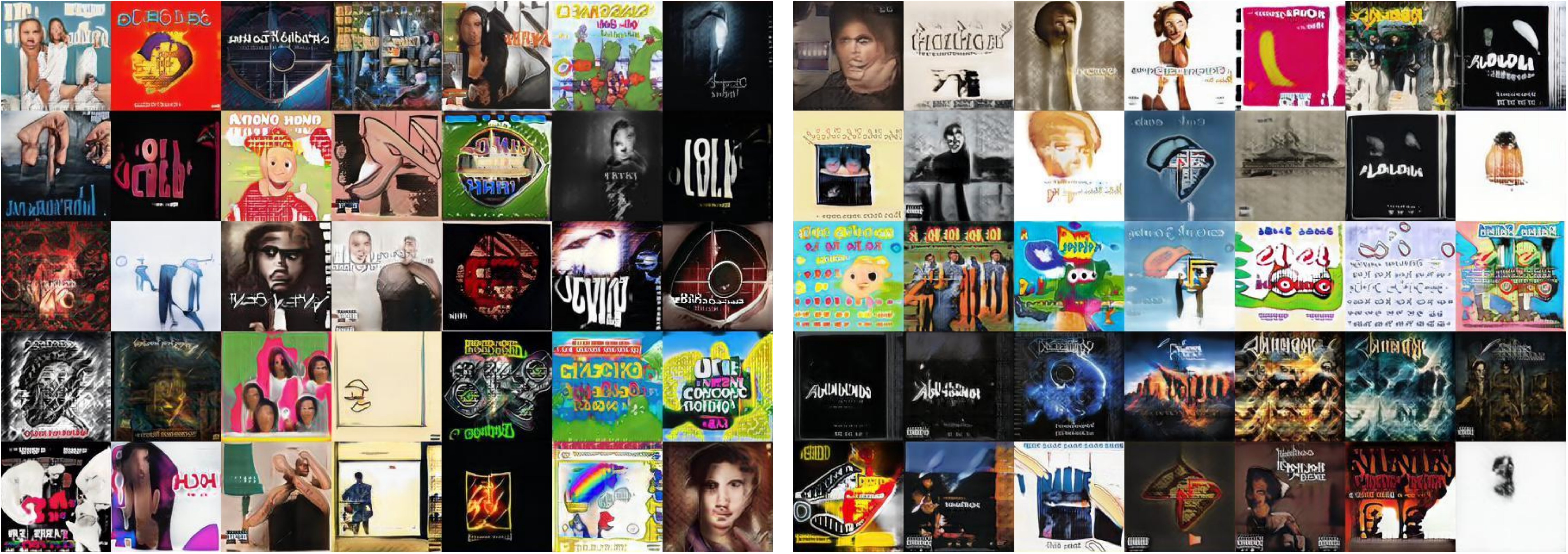}
    \caption{On the left: randomly generated, unconditional cover art. On the right: randomly generated cover art conditioned on genre (from top to bottom row: country, dance, kids, metal and rap).}
    \label{fig:StyleGAN2_results}
\end{figure*}

Our approach to generate audio-conditioned cover art differs from previous work in a number of aspects.
First, we eliminate the conditioning paths from the generator network by replacing it with a post-hoc, iterative, condition-driven optimization procedure.
With such a strategy, the generated image is gradually steered to comply with one or more fitness functions that incorporate the conditioning information.
This technique is becoming increasingly popular -- especially in the text-to-image synthesis domain -- since no additional training is required: any pretrained model can readily be used.
One of its most recent and powerful manifestations is VQGAN-CLIP \cite{crowson:2022}, and we specifically note its extensions to music-related modalities such as music-to-video \cite{jang:2022} and audio-to-description \cite{wu:2022} generation.
Second, all these works make use of gradient-descent-based optimization procedures to drive conditional image synthesis.
However, given the complexity of the task and its associated jagged loss landscape, this is suboptimal with respect to getting stuck in local minima and the potential of generating adversarial examples.
As a quick solution, the authors of VQGAN-CLIP increase the learning rate of Adam to $0.15$ -- as opposed to the typical value of $0.001$ -- presumably to overcome nearby local minima.
On the other hand, as suggested by Fernandes et al.~\cite{fernandes:2020}, genetic algorithms are much better suited to explore the latent space of GANs, and have e.g.~been used to generate Super Mario \cite{volz:2018, schrum:2020} and Doom \cite{Giacomello:2019} levels that comply with a set of fitness functions.
For this reason, our approach will also make use of genetic algorithms to drive cover art synthesis using an audio-based fitness function, and we will show that this method outperforms gradient descent on a number of qualitative and quantitative metrics.
Our results showcase the great potential and wide applicability of our framework to other and more advanced generative applications in the music domain.

The remainder of this paper is structured as follows.
In Section \ref{sec:methodology} we will discuss our methodology regarding the training of GANs for cover art generation, the design of an audio-based fitness function and the construction of an audio-guided optimization procedure based on genetic algorithms.
Our data gathering, selection and preprocessing strategies are outlined in Section \ref{sec:dataset}, after which we will give the results of our experiments and provide some generated examples in Section \ref{sec:results}.

\section{Methodology}
\label{sec:methodology}
\begin{figure}[t]
    \centering
    \includegraphics[width=0.9\linewidth, trim={0 1cm 0 0}, clip]{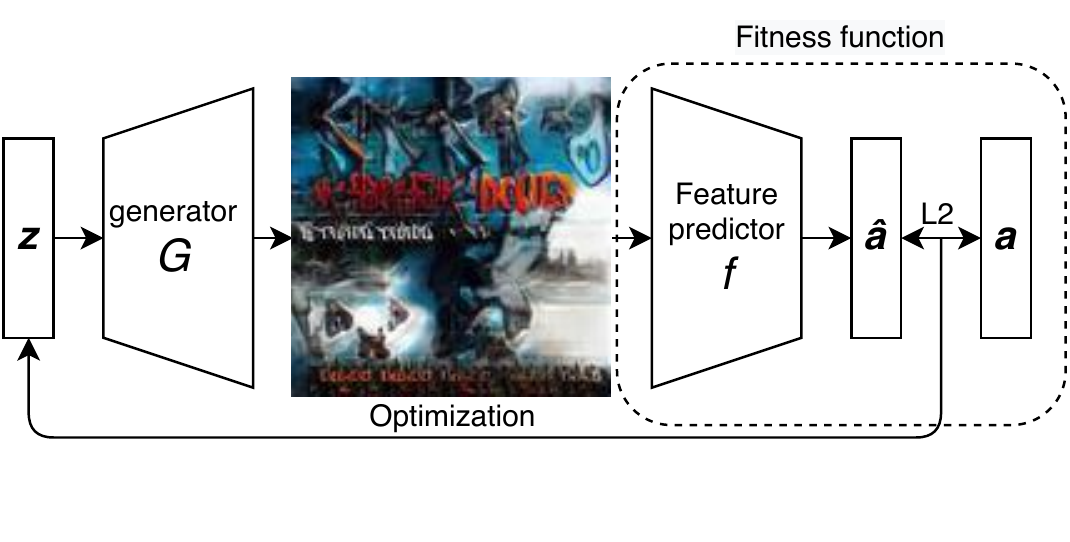}
    \caption{High-level overview of our proposed framework, where $\vect{z}$ is the random vector used to generate cover art, $\vect{a}$ are the target audio features and $\vect{\hat{a}}$ the predicted audio features. The feature predictor together with the $L_2$ distance between $\vect{a}$ and $\vect{\hat{a}}$ define the fitness function.}
    \label{fig:full_flow_inference}
\end{figure}

In order to generate cover art conditioned on audio, we propose a framework consisting of three main components: a cover art generator, an audio-based fitness function and an iterative optimization procedure.
As explained above, all components are trained separately and can, if needed, readily be replaced by other components with a similar function without the need to retrain the whole framework.
A high level overview is given in Figure \ref{fig:full_flow_inference}.
The flow starts by picking a song for which we want to generate cover art.
For this song we calculate its audio features $\vect{a}$.
A random vector $\vect{z}$ is then sampled to be used as input for the cover art generator $G$.
For a generated cover, a fitness function scores the match between this cover and the target audio features $\vect{a}$, and the optimization procedure will repeatedly update $\vect{z}$ and generate new covers in order to maximize the fitness score.
The cover with the maximum fitness score across the complete optimization procedure is chosen as the final result.
We will now look into the details of each individual component.

\subsection{Album Cover Art Generator}
\label{methodology:cover_generator}
The first component in our framework is the cover art generator, for which we train a state-of-the-art StyleGAN2 model $G$.
This model take a 512-dimensional standard normal vector $\vect{z}\!\sim\!\mathcal{N}_{512}(0,1)$ as input, and generates an RGB cover art image $G(\vect{z})$ with a resolution of $128\times 128$ pixels.
We choose to train two variants: an unconditional GAN and a GAN that is input-conditioned on a genre label.
This latter model will therefore be able to generate cover art for a predefined genre.
We will show in Section \ref{sec:results} that genre conditioning boosts the visual quality of the generated covers, but that it is not a necessary ingredient to achieve high fitness scores.


\subsection{Fitness function} \label{methodology:fitness_function}
The second component is a fitness function that scores how well a generated album cover matches with a target song.
As explained earlier, each song is represented in our framework by a set of audio features.
The fitness function will therefore need to bridge the gap between the visual and audio domain.
We opt for the following approach: we first train a neural network regression model $f_\theta$ that predicts the audio features belonging to a given ground-truth album cover as input.
Once this network is trained, we use it to predict the audio features of an album cover $G(\vect{z})$ generated from the latent vector $\vect{z}$.
The fitness score for this latent vector $\vect{z}$ is defined as the negative squared $L_2$ distance between the predicted audio features $f_\theta\!\left(G(\vect{z})\right)$ of the generated cover $G(\vect{z})$ and the ground truth audio features $\vect{a}$ for which we wish to generate an album cover:
\begin{align}
    \text{fitness}_{\vect{a}}(\vect{z}) \coloneqq - \lVert f_\theta\!\left(G(\vect{z})\right) - \vect{a} \rVert^2_2
\end{align}
It is clear that, in order to maximize the fitness, we need to minimize the $L_2$ distance.

\begin{figure}[t!]
    \centering
    \includegraphics[width=0.9\linewidth]{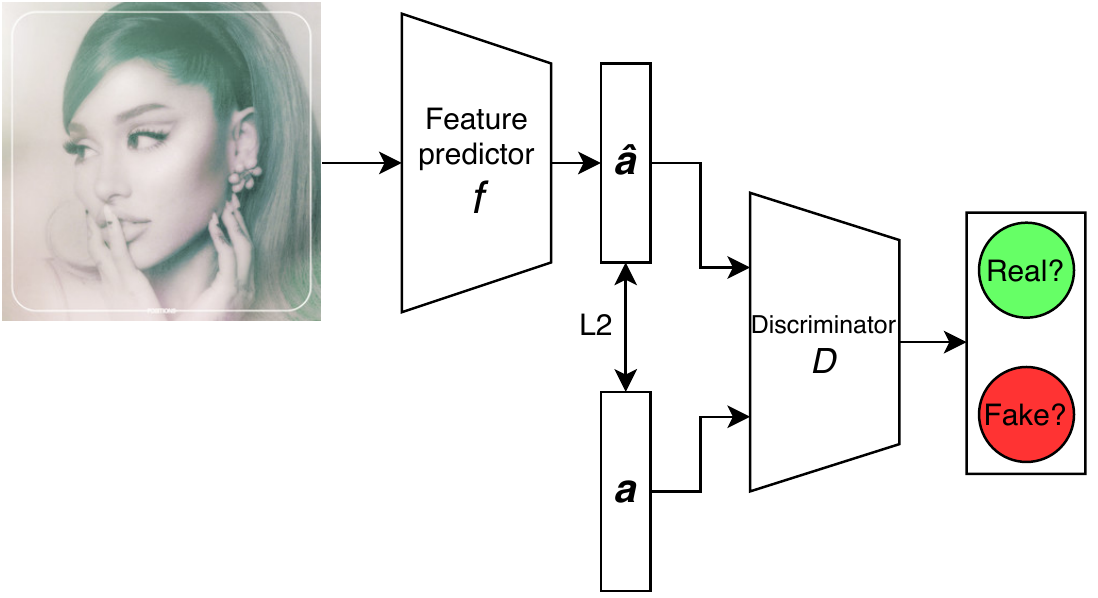}
    \caption{Architectural overview of the adversarial training procedure for the fitness function.}
    \label{fig:audio_feature_pred_architecture}
\end{figure}

To train the regression model, we perform supervised learning on a paired dataset $\mathcal{D} = \set{\left(\vect{c}, \vect{t}\right)}$ of album covers and audio features, where we use the latter as supervised targets.
We use the standard $L_2$ regression loss function, but we add an additional adversarial regularization term to force the predicted audio features to be close to realistic samples from the audio feature data space:
\begin{align}
    \min_\theta &\max_\phi\; \expectationwrt{(\vect{c},\vect{t})}{\lVert f_\theta(\vect{c}) - \vect{t}\rVert^2_2}\nonumber \\ +\, &\lambda\cdot\left( \expectationwrt{\vect{t}}{\log D_\phi(\vect{t})} +  \expectationwrt{\vect{c}}{\log\left(1 - D_\phi(f_\theta(\vect{c}))\right)} \right)
\end{align}
In the equation above, $D_\phi$ is a separate discriminator model that learns to predict whether an audio feature vector comes from the data set or is predicted by $f_\theta$.
It is then up to $f_\theta$ to fool $D_\phi$ as much as possible by outputting realistic audio feature predictions.
This is needed because preliminary experiments without adversarial regularization have shown that $f_\theta$ remains too cautious by predicting individual feature values close to the mode of their distribution.
This renders the entire audio feature vector unrealistic.
The adversarial regularization turns out to be a necessary ingredient to counter this phenomenon.
Figure \ref{fig:audio_feature_pred_architecture} visualizes the entire training flow.


\subsection{Optimization procedure}
\label{methodology:optimization_method}
The last component in our framework is an optimization procedure that steers the latent code $\vect{z}$ to maximize the fitness.
To perform this optimization, we discuss two different approaches: gradient descent and genetic algorithms.

As explained in the introduction, previous work mostly relies on \textbf{gradient descent} to optimize and explore the latent space of GANs \cite{crowson:2022, jang:2022, wu:2022}.
In such an approach, the parameters in both $G$ and $f_\theta$ are frozen, and the gradient of the fitness function w.r.t.~$\vect{z}$ is calculated through backpropagation \cite{bojanowski:2017,xiang:2017}.
To update $\vect{z}$ in the preferred direction, any gradient-descent-based optimizer, such as Adam, can be used.
The process is repeated for a number of iterations or until the fitness score reaches a certain threshold.
In spite of the simplicity of this method, one of its main issues is that it can easily get stuck in local minima or produce adversarial examples.
It is for this reason, presumably, that Crowson et al.~use a relatively large learning rate of $0.15$ compared to the default value of $0.001$ for Adam \cite{crowson:2022}.

As an alternative to gradient descient, we propose \textbf{genetic algorithms} (GA) \cite{whitley:1994,fernandes:2020}.
Genetic algorithms are based on the concept of evolution, where fitter individuals have a higher chance to procreate.
Commonly, three types of operations take place in a GA. 
In the \textit{mutatation} step, a set of genes is randomly adjusted, mimicking biological gene mutation.
In the \textit{crossover} step, information from individuals is interchanged, mimicking chromosomal crossover.
Finally, in the \textit{selection} step the fittest individuals are retained for the next generation.
In the context of this work, each individual is represented by a latent code $\vect{z}$.
At each new generation, we perform a series of crossover operations on two randomly selected individuals in the current pool.
The crossover point is sampled uniformly across the dimensionality of $\vect{z}$.
Next, we perform a random mutation on $m\%$ of the individuals in the pool by adding randomly sampled noise $\sim\mathcal{N}(0, 0.1)$ to their latent code.
Finally, for each $\vect{z}$ in the generation, we generate a cover and calculate its fitness score $\text{fitness}_{\vect{a}}(\vect{z})$ w.r.t.~the target audio features $\vect{a}$.
The $k\%$ fittest individuals are then selected to bootstrap the next generation.
As with the gradient descent method, this evolutionary process is repeated for a number of iterations or until the fitness score reaches a certain threshold.

\section{Dataset}
\label{sec:dataset}
We assemble a dataset of songs for which we collect the album cover, the audio features and the genre label.
Through the Spotify API we launch a series of textual search queries containing a year and a genre.
The year corresponds to the album release date -- which allows for a more diverse dataset -- and ranges between 1970 and 2022.
For the genres we choose between five diverse labels: country, dance, kids, metal and rap.
For each query we record the first 1000 returned tracks.
For each track, we collect the associated album covers which we downsize to $128\times 128$ pixels.
Spotify also returns 12 audio-focused features, and we retain the 9 features that are non-categorical\footnote{In preliminary experiments, the categorical features turned out to have little predictive power in genre classification, so we decided not to take them further into account. As a side-benefit, only dealing with continuous features allows for a simpler fitness function.}: danceability, valence, energy, tempo, loudness, speechiness, instrumentalness, liveness and acousticness.
These features are rescaled to the range $[0, 1]$.
Finally, the dataset is balanced across the five genres. 
Table \ref{tab:dataset} shows the complete dataset statistics\footnote{The dataset will be publicly released after acceptance.}.


\begin{table}[t!]
\begin{center}
\begin{tabular}{c|c c}
\toprule
& \textbf{Unique tracks} & \textbf{Unique album covers} \\
\hline
\textbf{Train} & $14705\times 5$ & $4675\times 5$ \\
\textbf{Test} & $2668\times 5$ & $822\times 5$ \\
\hline
\textbf{Total} & $\mathbf{17373\times 5}$ & $\mathbf{5497\times 5}$\\
\bottomrule
\end{tabular}
\caption{\label{tab:dataset}Dataset statistics: the numbers are shown per genre, and multiplied by $5$ to get the total dataset size across all five genres.}
\end{center}
\end{table}

\section{Experiments}
\label{sec:results}
In this section we will quantify visual quality, the influence of different optimization procedures and to what extent the genre of a song is contained in the generated cover.
We will also perform two qualitative experiments in which we examine the visual effect of changing a specific audio feature and to what extent the generated cover differs across tracks from the same album.

Each of our GAN models is trained for 364 epochs on the complete set of album covers in the dataset, which corresponds to 10 million image views during training.
All other parameters are fixed to their default StyleGAN2 values.
The audio feature predictor $f_\theta$ is bootstrapped with a pre-trained ResNet50 in which we changed the final layer to correspond with 9 continuous outputs, corresponding with each of the individual audio features.
The discriminator $D_\phi$ is a simple logistic regression model with 9 inputs and a single sigmoid output.
We empirically found a value of $\lambda = 9$ to give satisfactory results.
The complete fitness function is trained for $100$ epochs on the train set with Adam and the default learning rate $0.001$.
Regarding the genetic algorithm, we set the population size to 250, the mutation rate $m$ to $5\%$ and the selection threshold $k$ to $20\%$, which means that $80\%$ of the individuals in a generation are the result of crossover with the fittest individuals from the previous generation.
We always optimize for 200 iterations.

\subsection{Visual quality}
First, we will evaluate the image quality of the generated album covers by the conditional and unconditional GANs.
For this, we calculate the clean Fr\'echet Inception Distance (FID) as a quantitative visual metric \cite{parmar:2022, heusel:2017}.

We generate 27485 random album covers for the unconditional GAN, which equals the number of unique album covers used during training.
For the conditional GAN, we generate 5497 album covers for each of the five genres, again totaling 27485 covers.
We then calculate the FID for each individual genre, and also across the entire generated set of covers.
Table \ref{tab:StyleGAN_FID_scores} summarizes the results.
We see that the conditional GAN achieves much better visual quality, but we can also observe notable differences between genres, with kids performing best and country worst.
We hypothesize that the kids genre is indeed somewhat easier than the other genres because of the extensive use of clean-cut, geometrical shapes and saturated colors.

Figure \ref{fig:StyleGAN2_results} shows some samples of generated cover art for both GANs by sampling random latent codes $\vect{z}$.
In terms of subjective visual quality, the difference between both GANs seems small -- despite the gap in FID scores.
Overall, the generated covers have attractive colors and shapes, but still have a rather abstract, incoherent look.
This is something that can potentially be remedied by more powerful generative networks or by enlarging the dataset.
For the conditional GAN, each row on the right corresponds to resp.~country, dance, kids, metal and rap.
We can notice a clear visual distinction between each genre -- note, for example, the saturated colors of the kids cover, the dark dystopian mood of the metal covers, and the `parental advisory' logo on the rap covers.

\begin{table}[t!]
\begin{center}
\begin{tabular}{l|c} 
\toprule
& \textbf{clean FID} \\
\hline
\textbf{Unconditional} &   49.09 \\
\textbf{Conditional} &   35.39 \\
\hline
\textbf{Country} &  57.76 \\
\textbf{Dance} &   49.11 \\
\textbf{Kids} &   39.49 \\
\textbf{Metal} &    50.91 \\
\textbf{Rap} &    47.77 \\
\bottomrule
\end{tabular}
\caption{\label{tab:StyleGAN_FID_scores}Clean FID scores for the conditional and unconditional GANs, and for each individual genre using the conditional GAN (lower is better).}
\end{center}
\end{table}

\begin{figure*}[t]
    \centering
    \includegraphics[width=\linewidth]{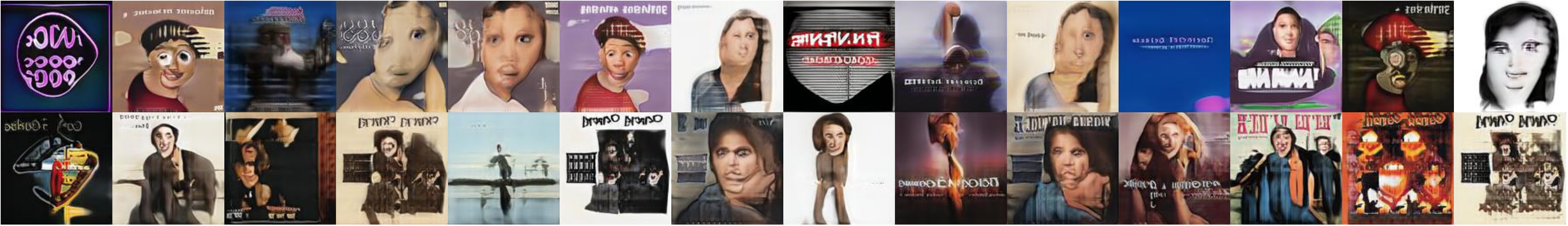}
    \caption{Optimized covers for multiple songs from a single country album. The first row and second row show resp.~the unconditional and conditional generator (conditioned on the country genre).}
    \label{fig:multiple_album_combined}
\end{figure*}

\begin{figure*}[t]
    \centering
    \includegraphics[width=\linewidth]{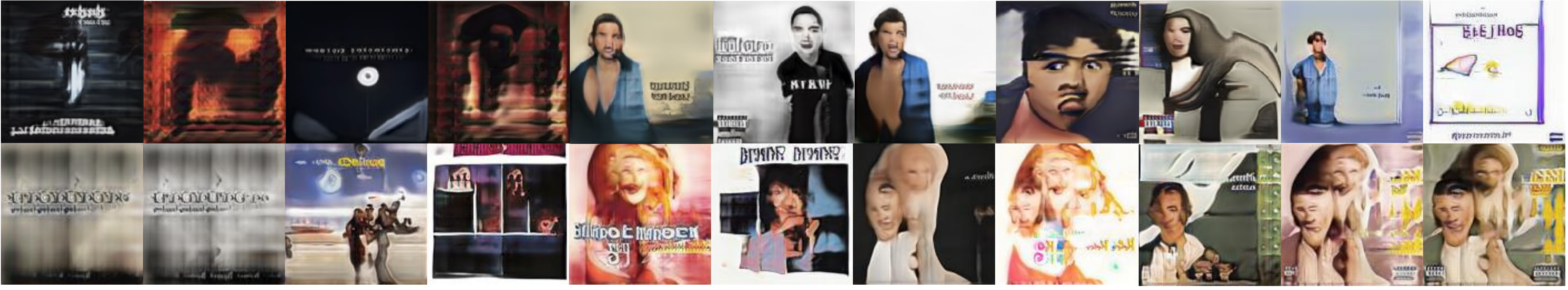}
    \caption{Optimized covers when interpolating the danceability feature between 0 (left) and 1 (right). The first row and second row show resp.~the unconditional and conditional generator (conditioned on the dance genre).}
    \label{fig:change_dimension_combined}
\end{figure*}

\subsection{Optimization effectiveness}
\label{sec:experiments:optimization}

To benchmark the optimization procedures, we generate cover art conditioned on 30 songs from the test set: six for each of the five genres. We test five different procedures: our proposed genetic algorithm (GA), and four gradient descent configurations: Adam and RMSprop, each with a large (0.15 -- as used in VQGAN-CLIP \cite{crowson:2022}) and small (0.001) learning rate.
We use 200 iterations for the genetic algorithm and 400 for the gradient descent methods (as used in VQGAN-CLIP).
Table \ref{tab:average_fitness} shows the average of obtained mean squared errors (MSE) between predicted and ground-truth audio features.
We notice that RMSprop tends to beat Adam, and that a larger learning rate indeed leads to an increased fitness score. But the genetic algorithm outperforms both RMSprop and Adam by a large margin.
We also observe little difference in performance between the two GANs when using GA.

\begin{figure*}[t!]
    \centering
    \includegraphics[width=\linewidth]{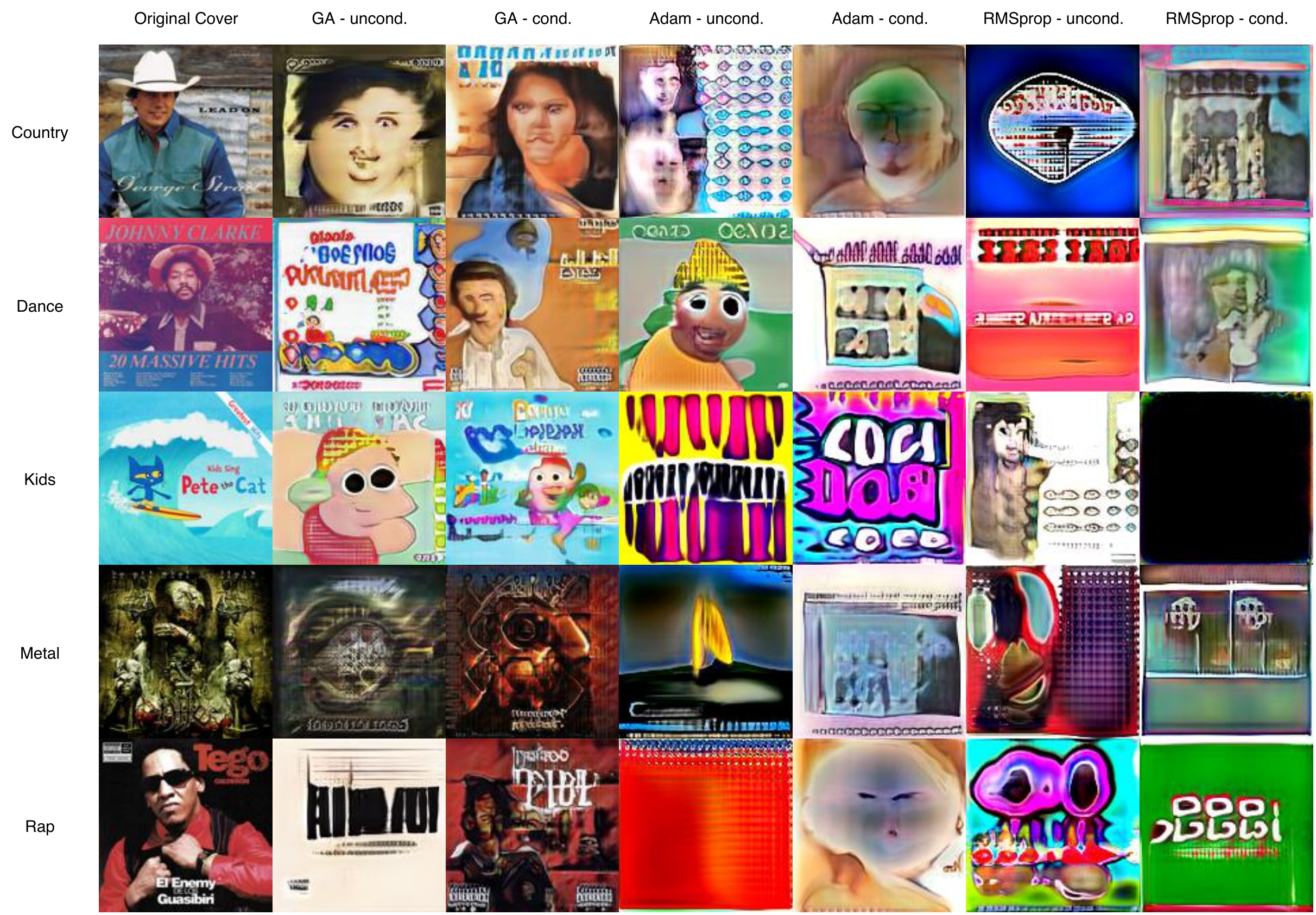}
    \caption{Optimized album covers for five tracks, one for each of the five genres (country, dance, kids, metal and rap). The left column shows the original album cover and the next columns show the effect of different optimization procedures.}
    \label{fig:results_optimization_combined}
\end{figure*}

Figure \ref{fig:results_optimization_combined} showcases generated covers using different optimizers.
Each row corresponds to a song in one of the five genres.
The first column is the original album cover, and the next columns represent the different optimization procedures.
The figure visually confirms our earlier conclusion that GA outperforms Adam and RMSprop (we only show learning rate 0.15).
We notice that Adam and RMSprop tend to collapse to incoherent color blobs.
The genetic algorithm, on the other hand, subjectively provides higher quality images with increased global and local structure.

\begin{table}[t!]
\begin{center}
\begin{tabular}{l|c c} 
\toprule
& \textbf{Unconditional} & \textbf{Conditional} \\
\hline
\textbf{GA} & \textbf{2.31} &  \textbf{2.37} \\
\textbf{Adam (0.15)} & 9.66 & 8.87 \\
\textbf{RMSprop (0.15)} & 8.92 & 5.72 \\
\textbf{Adam (0.001)} & 13.32 & 14.03 \\
\textbf{RMSprop (0.001)} & 11.59 & 14.42 \\
\bottomrule
\end{tabular}
\caption{\label{tab:average_fitness}Average MSE$\times 10^{-3}$ between the predicted and ground truth audio features for different optimizers.}
\end{center}
\end{table}

\subsection{Cover-based genre prediction}
To assess whether genre information is retained in the optimized cover art, we train a neural network classifier to predict the genre of a song based on its album cover.
For this, we use a pre-trained ResNet50 where we replace the last layer to output a probability distribution across the five genres.
We train this model on the train set for 100 epochs with Adam (learning rate 0.001), and it achieves an average accuracy of 58.83\% on the test set.

Table \ref{tab:genre_prediction_accuracy} shows the average accuracy of the genre prediction task across the 30 optimized album covers from Section \ref{sec:experiments:optimization}.
We test GA against and Adam and RMSprop with learning rate 0.15.
Once again, we can conclude that the genetic algorithm outperforms Adam and RMSprop, and both the conditional and unconditional GANs perform equally well.

\begin{table}[t!]
\begin{center}
\begin{tabular}{l|c c} 
\toprule
& \textbf{Unconditional} & \textbf{Conditional} \\
\hline
\textbf{GA} & \textbf{40\%} &  \textbf{40\%} \\
\textbf{Adam (0.15)} & 26.67\% &  30.0\% \\
\textbf{RMSprop (0.15)} & 26.67\% &  30.0\%  \\
\bottomrule
\end{tabular}
\caption{\label{tab:genre_prediction_accuracy}Average genre prediction accuracy on the optimized album covers from Section \ref{sec:experiments:optimization}.}
\end{center}
\end{table}

\subsection{Qualitative Experiments}
We conclude this paper with two qualitative experiments.
We only consider the GA optimization procedure, as it performed best in the previous experiments.

Up until now, we generated cover art for individual songs.
Despite some popular claims, the album concept is not dead and is slowly reinventing itself in the modern streaming era \cite{merlini:2020}, so it is valid to wonder if there would be any consistency between generated covers for songs that appear on the same album.
We choose an arbitrary country album\footnote{Live Like You Were Dying - Tim McGraw} from the test set and generate a cover with both GANs for each of the 14 songs on this album.
Figure \ref{fig:multiple_album_combined} shows the results of this experiment.
We notice that some of the generated covers look more or less similar -- with a face-like figure in most of them -- but some look significantly different.
This is, of course, to be expected since an album can consist of a diverse set of songs.



In a final experiment, we want to investigate the visual effect of changing a single dimension in the audio feature vector.
We take an arbitrary dance song\footnote{34+35 - Ariana Grande} from the test set and choose to change the value of the danceability dimension (original value 0.83) while keeping the other dimensions fixed.
More specifically, we vary the danceability feature between 0 and 1 in steps of 0.1.
The results of this experiment for both GANs can be seen in Figure \ref{fig:change_dimension_combined}.
For low danceability, we notice significantly darker or less saturated covers, and when the danceability is increased, the covers get more expressive and colorful.
Although this is a confined, limited and subjective experiment, we can carefully conclude that our framework indeed enforces the cover art generator to lock onto specific audio features and to reflect these changes in the generated covers.



\section{Conclusion}
In this paper we tackled the task of generating album cover art conditioned on audio information.
We proposed a flexible framework in which each component can be trained individually and can be interchanged with other components without the need to retrain the complete framework.
We presented a novel approach to model the fitness between cover art and audio features, by combining supervised training with an adversarial side-objective.
Finally, we showed that genetic algorithms outperform gradient-descent-based approaches to optimizing and exploring the latent space of generative adversarial networks, and that they are the preferred method for conditional image synthesis tasks.
As a downside, the generated cover art presented in this work is far from production-ready -- reduced resolution and lack of local detail and definition -- but thanks to the flexibility of our framework, a more powerful generative network that might potentially solve these issues can readily be used.
But for now, the generated covers can certainly be used as a source of creative inspiration.

We believe that this work paves the road for many extensions and more advanced applications in the audiovisual generative domain.
One extension is the application of our model to albums instead of individual songs, as was briefly touched upon in the experiments.
Evidently, if we want to design cover art for a new album, we want one cover that fits the album as a whole.
Next to conditioning on multiple songs, we might also think of other data modalities to condition on, e.g. the album name, the artist name, the artist's face, overall color scheme, the presence of certain objects, other album covers to be used as inspiration, etc.

\clearpage
\bibliography{lib}

\end{document}